# Hybrid Joint Diagonalization Algorithms

Mohamed Nait-Meziane, Karim Abed-Meraim, Abd-Krim Seghouane, and Ammar Mesloub

*Abstract*—This paper deals with a hybrid joint diagonalization (JD) problem considering both Hermitian and transpose congruences. Such problem can be encountered in certain non-circular signal analysis applications including blind source separation. We introduce new Jacobi-like algorithms using Givens or a combination of Givens and hyperbolic rotations. These algorithms are compared with state-of-the-art methods and their performance gain, especially in the high dimensional case, is assessed through simulation experiments including examples related to blind separation of non-circular sources.

## I. Introduction

JOINT decomposition of matrices sharing the same algebraic structure is an important problem with many engineering applications. In particular, the joint diagonalization (JD) problem is found in several signal processing applications related to blind source separation (BSS) [1], to multidimensional parameter estimation and pairing [2], to blind system identification [3], and to tensor decomposition [4].

Different types of JD problems exist including the algebraic one where the considered set of matrices is of the form $\{\mathbf{M}_k = \mathbf{A}\mathbf{D}_k\mathbf{A}^{-1}\}_{1 \leq k \leq K}$, $\mathbf{A}$ being a non-singular matrix and $\mathbf{D}_k$ are diagonal matrices [5]. Common JD problems include the JD by Hermitian congruence [1] where matrices $\mathbf{M}_k, k = 1, \ldots, K$ share the common structure $\mathbf{M}_k = \mathbf{A}\mathbf{D}_k\mathbf{A}^H$; the JD by transpose congruence where $\mathbf{M}_k = \mathbf{A}\mathbf{D}_k\mathbf{A}^T, k = 1, \ldots, K$; and the hybrid JD (HJD) where two sets of complex matrices $\{\mathbf{M}_k = \mathbf{A}\mathbf{D}_k\mathbf{A}^H\}_{1 \leq k \leq K_1}$ and $\{\mathbf{N}_k = \mathbf{A}\mathbf{L}_k\mathbf{A}^T\}_{1 \leq k \leq K_2}$ ($\mathbf{L}_k$ being diagonal matrices) are considered.

It is the latter case that is treated in this paper. It is mostly encountered when dealing with the statistics of multivariate non-circular complex data (see for example [6]). Among the existing solutions for this HJD problem one can cite the extended version of FAJD proposed in [7], the NOODLES algorithm proposed in [8], which relies on a natural-gradient technique, the algorithm proposed in [9] based on weighted least-squares (WLS) criterion, and the alternating least squares (ALS) algorithm proposed in [10], which considers $\mathbf{A}^H$ and $\mathbf{A}^T$ as different variables during the iterative process. All of these methods consider the non-orthogonal case where the matrix $\mathbf{A}$ is non-unitary. However, in the context of blind source separation, one might first apply a data whitening, which renders matrix $\mathbf{A}$ unitary, before fully estimating it using orthogonal JD (see for example SOBI algorithm in [1]).

M. Nait-Meziane and K. Abed-Meraim are with the PRISME Laboratory, University of Orléans, 12 rue de Blois, 45067 Orléans, France.
Abd-Krim Seghouane is with the Department of Electrical and Electronic Engineering, University of Melbourne, Parkville, Australia.
Ammar Mesloub is with the Laboratoire Traitement du signal, École Militaire Polytechnique, Algiers, Algeria.

This case is considered briefly in [11]. In this paper, we start by dealing with the orthogonal case and introduce several extensions of the work in [11]. For the general non-orthogonal case, it has been shown recently that many of the standard JD methods fail to achieve good JD performance in adverse scenarios (see [12] for details). In such cases, the CJDi algorithm introduced in [12], seems to be one of the most robust and effective methods for the standard JD by Hermitian congruence. Hence, we introduce an extended version of CJDi (named H-CJDi) to deal with the HJD problem in the non-orthogonal situation. The effectiveness of the proposed algorithms is discussed and illustrated through simulation results and an application example of blind separation of non-circular sources.

## II. Problem Formulation and Basic Concepts

Consider two sets of $n \times n$ complex matrices satisfying

$$\mathbf{M}_k = \mathbf{A}\mathbf{D}_k\mathbf{A}^H, \quad k = 1, \ldots, K_1 \quad (1)$$

$$\mathbf{N}_k = \mathbf{A}\mathbf{L}_k\mathbf{A}^T, \quad k = 1, \ldots, K_2 \quad (2)$$

where $\mathbf{D}_k \in \mathbb{C}^{n \times n}$ and $\mathbf{L}_k \in \mathbb{C}^{n \times n}$ are diagonal matrices, and $\mathbf{A} \in \mathbb{C}^{n \times n}$ is an unknown matrix (called mixing matrix in the BSS context). The JD problem consists of finding a matrix $\mathbf{V} \in \mathbb{C}^{n \times n}$ such that matrices $\mathbf{V}^H\mathbf{M}_k\mathbf{V}, k = 1, \ldots, K_1$ and $\mathbf{V}^H\mathbf{N}_k\mathbf{V}^*, k = 1, \ldots, K_2$ ($\mathbf{V}^*$ being the complex conjugate of $\mathbf{V}$) are diagonal. For the approximate JD problem, an additive error term affects matrices $\mathbf{M}_k$ and $\mathbf{N}_k$ in which case $\mathbf{V}$ is sought in such a way it minimizes the following function

$$\mathcal{S}(\mathbf{V}) = \sum_{k=1}^{K_1} \text{off}(\mathbf{V}^H\mathbf{M}_k\mathbf{V}) + \sum_{k=1}^{K_2} \text{off}(\mathbf{V}^H\mathbf{N}_k\mathbf{V}^*) \quad (3)$$

where $\text{off}(\mathbf{X}) = \sum_{1 \leq i \neq j \leq n} |X_{ij}|^2$. To solve this problem, we consider two standard situations: the one where matrix $\mathbf{A}$ is orthogonal (i.e. $\mathbf{A}^H\mathbf{A} = \mathbf{I}$) corresponding to the case where a data pre-whitening is applied (see [1] for details) and the general case of non-orthogonal matrix $\mathbf{A}$ used when the pre-whitening is not possible or poorly achieved (due for example to a short sample size [13]). The two cases are considered in the sequel where criterion (3) is iteratively optimized through Givens (or a combination of Givens and hyperbolic) rotations.

## III. Hybrid Joint Diagonalization Algorithms

### A. Orthogonal case

*1) Complex Orthogonal HJD (CO-HJD) algorithm:* Here, $\mathbf{V}$ is decomposed as a product of Givens rotations

$$\mathbf{V} = \prod_{\#\text{sweeps}} \prod_{1 \leq p < q \leq n} \mathbf{G}_{pq}(\theta, \alpha) \quad (4)$$

where #sweeps is the number of iterations and $\mathbf{G}_{pq}(\theta,\alpha)$ is equal to the identity matrix except for its $(p,p)^{\text{th}}$, $(p,q)^{\text{th}}$, $(q,p)^{\text{th}}$, and $(q,q)^{\text{th}}$ entries given by

$$\begin{bmatrix} G_{pp} & G_{pq} \\ G_{qp} & G_{qq} \end{bmatrix} = \begin{bmatrix} \cos(\theta) & -\sin(\theta)e^{-j\alpha} \\ \sin(\theta)e^{j\alpha} & \cos(\theta) \end{bmatrix} \quad (5)$$

The minimization of criterion (3) with respect to $\mathbf{G}_{pq}(\theta,\alpha)$ is equivalent to the following optimization problem[1] [11]:

$$\max_{\mathbf{v}} \mathbf{v}^T \operatorname{Re}\left(\mathbf{E}_1^H \mathbf{E}_1 - \mathbf{E}_2^H \mathbf{E}_2\right) \mathbf{v} \quad \text{s.t.} \quad \mathbf{v}^T \mathbf{v} = 1, \quad (6)$$

where $\mathbf{v} = [\cos(2\theta), -\sin(2\theta)\cos(\alpha), -\sin(2\theta)\sin(\alpha)]^T$, $\operatorname{Re}(\cdot)$ is the real-part operator, $\mathbf{E}_1^T = [\mathbf{e}_{1,1},\ldots,\mathbf{e}_{1,K_1}]$ with $\mathbf{e}_{1,k} = [M_{k,pp} - M_{k,qq}, -(M_{k,pq} + M_{k,qp}), j(M_{k,qp} - M_{k,pq})]^T$, and $\mathbf{E}_2^T = [\mathbf{e}_{2,1},\ldots,\mathbf{e}_{2,K_2}]$ with $\mathbf{e}_{2,k} = [2N_{k,pq}, N_{k,pp} - N_{k,qq}, j(N_{k,pp} + N_{k,qq})]^T$ [2]. The solution of (6) is the principal eigenvector of this quadratic form matrix.

*2) Real Orthogonal HJD (RO-HJD) algorithm:* In [11], the authors pointed out the possibility to estimate $\mathbf{A}$, after prewhitening, up to an unknown real orthogonal matrix[3]. The latter can be estimated in a second stage by using real Givens rotations. Considering a unitary matrix $\mathbf{A}$ and assuming the matrix $\mathbf{N}_1$ is full rank, then we can transform $\mathbf{A}$ into an orthogonal real-valued matrix thanks to the following result.

**Lemma 1.** *Let $\mathbf{U}$ be the matrix of left singular vectors of $\mathbf{N}_1$ and consider the eigendecomposition of $\mathbf{U}^H \mathbf{N}_1 \mathbf{U}^* = \mathbf{E}\mathbf{S}\mathbf{E}^T$, with $\mathbf{S} = \operatorname{diag}(r_1 e^{j2\alpha_1},\ldots,r_n e^{j2\alpha_n})$ then $\mathbf{B} = \mathbf{U}\mathbf{E}\tilde{\mathbf{S}}$, $\tilde{\mathbf{S}} = \operatorname{diag}(e^{j\alpha_1},\ldots,e^{j\alpha_n})$, is equal to $\mathbf{A}$ up to an unknown real orthogonal matrix $\mathbf{Q}$, i.e. $\mathbf{B}^H \mathbf{A} = \mathbf{Q}$.*

Since matrix $\mathbf{B}$ transforms $\mathbf{A}$ into a real orthogonal matrix, then RO-HJD method consists of applying the standard JD algorithm in [14] using real Givens rotations (with $\alpha = 0$) to the transformed matrices $\mathbf{B}^H \mathbf{M}_k \mathbf{B}$ and $\mathbf{B}^H \mathbf{N}_k \mathbf{B}^*$.

*3) Augmented Real Orthogonal HJD (ARO-HJD) algorithm:* Here, we present a way of solving the HJD problem in the real domain using the statistics (e.g., correlation matrices) of the augmented real vector [15] $\overline{\mathbf{x}}(t) = [\operatorname{Re}(\mathbf{x}(t))^T, \operatorname{Im}(\mathbf{x}(t))^T]^T$ ($\operatorname{Im}(\cdot)$ being the imaginary-part operator). When the observed vector corresponds to the mixtures of $n$ independent sources, i.e. $\mathbf{x}(t) = \mathbf{A}\mathbf{s}(t), t = 1,\ldots,T$, the statistics of $\overline{\mathbf{x}}(t)$ would have the form in (2) where the diagonal matrix is replaced by a 4-diagonal-blocks matrix (since the real and imaginary parts of the source signals are not necessarily independent). In other words, the HJD becomes equivalent to a block JD (BJD) problem in the real domain. By minimizing an appropriate BJD criterion (chosen in such a way one targets the diagonal-blocks structure of the considered matrices) one can achieve the desired matrix decomposition with a Jacobi-like algorithm using real Givens rotations. However, we found that the resulting algorithm has a convergence rate slighlty lower than those of CO-HJD and RO-HJD algorithms, and so due to space limitation its derivation details are omitted.

Instead, we present here a specific case where the source signals are non-circular but their real and imaginary parts are independent and need to be separated which is a key problem in certain BSS applications (e.g., [16], [17]). In that case, the statistics of $\overline{\mathbf{x}}(t)$ have the form $\overline{\mathbf{M}}_k = \overline{\mathbf{A}}\mathbf{D}_k \overline{\mathbf{A}}^T$ where $\mathbf{D}_k$, $k = 1,\ldots,K$ are diagonal matrices and

$$\overline{\mathbf{A}} = \begin{bmatrix} \operatorname{Re}(\mathbf{A}) & -\operatorname{Im}(\mathbf{A}) \\ \operatorname{Im}(\mathbf{A}) & \operatorname{Re}(\mathbf{A}) \end{bmatrix}. \quad (7)$$

We propose to minimize (3) using real Givens rotations $\mathbf{G}_{p,q}(\theta)$. However, to preserve the special structure of $\overline{\mathbf{A}}$, one needs to simultaneously apply Givens rotations $\mathbf{G}_{p,q}(\theta)$ and $\mathbf{G}_{p+n,q+n}(\theta)$ and also simultaneously apply $\mathbf{G}_{p,q+n}(\theta')$ and $\mathbf{G}_{q,p+n}(\theta')$ (for more details refer to [12]). The unmixing matrix is defined as

$$\overline{\mathbf{V}} = \prod_{\text{\#sweeps}} \prod_{p=1}^{n} \mathbf{G}_{p,p+n}(\theta'') \prod_{q=p+1}^{n} \overline{\mathbf{V}}'_{pq}(\theta')\overline{\mathbf{V}}_{pq}(\theta) \quad (8)$$

with $\overline{\mathbf{V}}_{pq}(\theta) = \mathbf{G}_{p,q}(\theta)\mathbf{G}_{p+n,q+n}(\theta)$ and $\overline{\mathbf{V}}'_{pq}(\theta') = \mathbf{G}_{p,q+n}(\theta')\mathbf{G}_{q,p+n}(\theta')$. Minimizing (3) w.r.t. $\theta$ leads to a quadratic form maximization, $\max_{\overline{\mathbf{v}}} \overline{\mathbf{v}}^T \mathbf{Q} \overline{\mathbf{v}}$ with $\overline{\mathbf{v}} = [\cos(2\theta), -\sin(2\theta)]^T$ and $\mathbf{Q} = \mathbf{F}_1^T \mathbf{F}_1 + \mathbf{F}_2^T \mathbf{F}_2$. For $\theta'$, the quadratic form becomes $\mathbf{Q}' = \mathbf{F}_3^T \mathbf{F}_3 + \mathbf{F}_4^T \mathbf{F}_4$, and for $\theta''$ [4], the quadratic form is $\mathbf{Q}'' = \mathbf{F}_5^T \mathbf{F}_5$, where $\mathbf{F}_i$, $i = 1,\ldots,5$ are defined as $\mathbf{F}_i^T = [\mathbf{f}_{i,1},\ldots,\mathbf{f}_{i,K}]$ with

$\mathbf{f}_{1,k} = [\overline{M}_{k,q,q} - \overline{M}_{k,p,p}, \overline{M}_{k,p,q} + \overline{M}_{k,q,p}]^T$,
$\mathbf{f}_{2,k} = [\overline{M}_{k,q+n,q+n} - \overline{M}_{k,p+n,p+n}, \overline{M}_{k,p+n,q+n} + \overline{M}_{k,q+n,p+n}]^T$,
$\mathbf{f}_{3,k} = [\overline{M}_{k,q+n,q+n} - \overline{M}_{k,p,p}, \overline{M}_{k,p,q+n} + \overline{M}_{k,q+n,p}]^T$,
$\mathbf{f}_{4,k} = [\overline{M}_{k,p+n,p+n} - \overline{M}_{k,q,q}, \overline{M}_{k,q,p+n} + \overline{M}_{k,p+n,q}]^T$,
$\mathbf{f}_{5,k} = [\overline{M}_{k,p+n,p+n} - \overline{M}_{k,p,p}, \overline{M}_{k,p,p+n} + \overline{M}_{k,p+n,p}]^T$.

The solutions are the unit-norm principal eigenvectors of matrices $\mathbf{Q}$, $\mathbf{Q}'$ and $\mathbf{Q}''$, respectively.

**Remark**: CO-HJD is the 'natural' extension of SOBI to the HJD case. RO-HJD is of interest only when the sources are strongly non-circular where it may help improve the JD quality (Fig. 3b) and ARO-HJD is useful when the separation of the real and imaginary components of the sources is required.

*B. Non-orthogonal case*

In this case, matrix $\mathbf{V}$ is decomposed as a product of Givens and hyperbolic rotations

$$\mathbf{V} = \prod_{\text{\#sweeps}} \prod_{1 \leq p < q \leq n} \mathbf{R}_{pq}(\theta,\alpha,y,\phi) \quad (9)$$

where $\mathbf{R}_{pq}(\theta,\alpha,y,\phi)$ is the elementary matrix, combining a Givens and a hyperbolic rotation, given by

$$\mathbf{R}_{pq}(\theta,\alpha,y,\phi) = \mathbf{G}_{pq}(\theta,\alpha)\mathbf{H}_{pq}(y,\phi) \quad (10)$$

where $\mathbf{G}_{pq}(\theta,\alpha)$ is defined in (5) and the hyperbolic rotation $\mathbf{H}_{pq}(y,\phi)$ is equal to the identity except for its $(p,p)^{\text{th}}$, $(p,q)^{\text{th}}$, $(q,p)^{\text{th}}$, and $(q,q)^{\text{th}}$ entries given by

$$\begin{bmatrix} H_{pp} & H_{pq} \\ H_{qp} & H_{qq} \end{bmatrix} = \begin{bmatrix} \cosh(y) & \sinh(y)e^{-j\phi} \\ \sinh(y)e^{j\phi} & \cosh(y) \end{bmatrix}. \quad (11)$$

---

[1] This algorithm's version was introduced in [11] but without any validation or numerical performance assessment.

[2] Entry $(i,j)$ of matrix $\mathbf{X}_k$ is written $X_{k,ij}$ or $X_{k,i,j}$.

[3] This claim was mentioned in [11] but without giving the proper way to achieve it.

[4] This rotation is introduced to remove the inherent phase indeterminacy of the BSS that might mix the real and imaginary signal components.

To find the optimal matrix $\mathbf{R}_{pq}$, the direct use of criterion (3) leads to a non-linear optimization problem with no closed-form solution. Hence, in [12] a simplified criterion was considered for the JD of complex Hermitian matrices consisting of minimizing at each step the sum of square modulus of the $(p,q)^{th}$ and $(q,p)^{th}$ entries of the transformed matrices (CJDi algorithm). In particular, it was shown that applying one matrix $\mathbf{R}_{pq}(\theta, \alpha, y, \phi)$ with a direct optimization of the JD criterion w.r.t. parameters $(\theta, \alpha, y, \phi)$ (which has no closed-form solution) can be replaced by applying two successive matrices $\mathbf{R}_{pq}^{(0)} = \mathbf{R}_{pq}(\theta, 0, y, 0)$ and $\mathbf{R}_{pq}^{(\frac{\pi}{2})} = \mathbf{R}_{pq}(\theta', \frac{\pi}{2}, y', \frac{\pi}{2})$, which has the advantage of closed-form solutions for the optimal pairs of parameters $(\theta, y)$ and $(\theta', y')$. Motivated by the effectiveness of the CJDi algorithm especially in the adverse scenarios (see [12] for details), we propose to generalize it for solving our hybrid JD problem. As in [12], we proceed by first transforming the $K_1$ matrices $\mathbf{M}_k$ into $2K_1$ Hermitian matrices $\{\tilde{\mathbf{M}}_k\}_{1 \leq k \leq 2K_1}$ such that (for $k = 1, \ldots, K_1$) $\tilde{\mathbf{M}}_{2k-1} = (\mathbf{M}_k + \mathbf{M}_k^H)/2$ and $\tilde{\mathbf{M}}_{2k} = (\mathbf{M}_k - \mathbf{M}_k^H)/(2j)$. Then, at each iteration and for each pair $(p,q)_{1 \leq p < q \leq n}$, we search successively for matrices $\mathbf{R}_{pq}^{(0)}$ and $\mathbf{R}_{pq}^{(\frac{\pi}{2})}$ minimizing, respectively, $\mathcal{C}(\mathbf{R}_{pq}^{(0)})$ and $\mathcal{C}(\mathbf{R}_{pq}^{(\frac{\pi}{2})})$ with

$$\mathcal{C}(\mathbf{V}) = \sum_{k=1}^{2K_1} |[\mathbf{V}^H \tilde{\mathbf{M}}_k \mathbf{V}]_{pq}|^2 + \sum_{k=1}^{K_2} |[\mathbf{V}^H \mathbf{N}_k \mathbf{V}^*]_{pq}|^2. \quad (12)$$

The minimization of the previous cost functions can be written, respectively, as (see supplementary material [18])[5]:

$$\min_{\mathbf{w}} \mathbf{w}^T \mathrm{Re}(\mathbf{E}_3^H \mathbf{E}_3 + \mathbf{E}_4^H \mathbf{E}_4) \mathbf{w} \quad \text{s.t.} \quad \mathbf{w}^T \mathbf{J}\mathbf{w} = 1, \quad (13)$$

$$\min_{\mathbf{w}'} \mathbf{w}'^T \mathrm{Re}(\mathbf{E}_5^H \mathbf{E}_5 + \mathbf{E}_6^H \mathbf{E}_6) \mathbf{w}' \quad \text{s.t.} \quad \mathbf{w}'^T \mathbf{J}\mathbf{w}' = 1 \quad (14)$$

where $\mathbf{w} = [\sinh(2y), -\sin(2\theta)\cosh(2y), \cos(2\theta)\cosh(2y)]^T$, $\mathbf{w}' = [\sinh(2y'), -\sin(2\theta')\cosh(2y'), \cos(2\theta')\cosh(2y')]^T$, $\mathbf{J} = \mathrm{diag}([-1, 1, 1])$, and matrices $\mathbf{E}_i$, $i = 3, \ldots, 6$ are defined as $\mathbf{E}_i^T = [\mathbf{e}_{i,1}, \ldots, \mathbf{e}_{i,2K_1}]$ for $i = 3, 5$ and $\mathbf{E}_i^T = [\mathbf{e}_{i,1}, \ldots, \mathbf{e}_{i,K_2}]$ for $i = 4, 6$ with $\mathbf{e}_{3,k} = [\tilde{M}_{k,pp} + \tilde{M}_{k,qq}, \tilde{M}_{k,pp} - \tilde{M}_{k,qq}, 2\mathrm{Re}(\tilde{M}_{k,pq})]^T$, $\mathbf{e}_{4,k} = [N_{k,pp} + N_{k,qq}, N_{k,pp} - N_{k,qq}, 2N_{k,pq}]^T$, $\mathbf{e}_{5,k} = [-(\tilde{M}_{k,pp} + \tilde{M}_{k,qq}), \tilde{M}_{k,qq} - \tilde{M}_{k,pp}, 2\mathrm{Im}(\tilde{M}_{k,pq})]^T$, and $\mathbf{e}_{6,k} = [N_{k,pp} - N_{k,qq}, N_{k,pp} + N_{k,qq}, -2jN_{k,pq}]^T$.

The optimal solution of (13) (resp. (14)) is the generalized eigenvector of median (smallest non-negative) generalized eigenvalue of $(\mathrm{Re}(\mathbf{E}_3^H \mathbf{E}_3 + \mathbf{E}_4^H \mathbf{E}_4), \mathbf{J})$ (resp. $(\mathrm{Re}(\mathbf{E}_5^H \mathbf{E}_5 + \mathbf{E}_6^H \mathbf{E}_6), \mathbf{J}))$ [12], [19]. This solution is normalized such that it satisfies the required optimization constraint.

## IV. SIMULATION RESULTS

We compare our proposed algorithms with the following ones: Second-Order Blind Identification (SOBI) [1], Fast Approximate Joint Diagonalization (FAJD) [20], Hybrid FAJD (H-FAJD) [7], NOODLES [8], and H-NOODLES [8][6]. Par-ticularly, we simulate cases representing adverse conditions under which JD can be difficult. For example, ill-conditioned mixing matrix $\mathbf{A}$, noisy target matrices, large dimensional target matrices, and non-unique JD condition[7]. The Modulus of Uniqueness (MoU) is an indicator of the uniqueness of the JD and is defined as [21] $\mathrm{MoU} = \max_{i,j} \left( \frac{|\mathbf{d}_i^H \mathbf{d}_j|}{||\mathbf{d}_i||\,||\mathbf{d}_j||} \right)$, $1 \leq i \neq j \leq n$ where $\mathbf{d}_i = [D_{1,ii}, \ldots, D_{K_1,ii}, L_{1,ii}, \ldots, L_{K_2,ii}]^T$. The JD quality decreases as MoU approaches 1.

---
[5]Contains mathematical derivation of CO-HJD, ARO-HJD and H-CJDi.
[6]For consistency, we changed the algorithms' names so that all hybrid versions are denoted with an "H". Originally, H-FAJD is termed ncSOBI or extended FAJD [7] whereas NOODLES is termed H-NOODLES (for Hermitian NOODLES; using only Hermitian matrices) and H-NOODLES is termed NOODLES (the hybrid version).

The noisy target matrices are modeled as

$$\mathbf{M}_k = \mathbf{A}\mathbf{D}_k \mathbf{A}^H + \mathbf{W}_k, \quad k = 1, \ldots, K_1 \quad (15)$$

$$\mathbf{N}_k = \mathbf{A}\mathbf{L}_k \mathbf{A}^T + \mathbf{W}'_k, \quad k = 1, \ldots, K_2 \quad (16)$$

where $\mathbf{W}_k$ and $\mathbf{W}'_k$ are perturbation matrices such that $\mathbf{W}_k = \delta_k \mathbf{B}_k$ (resp. $\mathbf{W}'_k = \delta'_k \mathbf{B}'_k$) where $\mathbf{B}_k$ (resp. $\mathbf{B}'_k$) is a random matrix generated with i.i.d. unit-variance complex Gaussian entries. The positive scalar $\delta_k$ (resp. $\delta'_k$) is tuned to achieve the desired Signal-to-Noise Ratio (SNR) defined for $\mathbf{M}_k$ (and similarly for $\mathbf{N}_k$) as $\mathrm{SNR(dB)} = 10\log_{10}\left(\frac{||\mathbf{A}\mathbf{D}_k \mathbf{A}^H||_F}{||\mathbf{W}_k||_F}\right)$. To evaluate and compare the JD performance, we use the following classical performance index (PI) [22]

$$\mathrm{PI}(\mathbf{P}) = \frac{1}{2n(n-1)} \sum_{l=1}^{n} \left( \sum_{m=1}^{n} \frac{|P_{lm}|^2}{\max_k |P_{lk}|^2} - 1 \right)$$

$$+ \frac{1}{2n(n-1)} \sum_{m=1}^{n} \left( \sum_{l=1}^{n} \frac{|P_{lm}|^2}{\max_k |P_{km}|^2} - 1 \right) \quad (17)$$

where $\mathbf{P} = \mathbf{V}^H \mathbf{A}$. Each point in the plots is a median value computed over 100 (resp. 20) Monte Carlo runs for small dimensional matrices of size $n = 5$ (resp. large dimensional matrices of size $n = 50$). We chose $K_1 = K_2 = 5$ matrices. Matrix $\mathbf{A}$ is generated randomly at each run with i.i.d. Gaussian entries (but with controlled condition value when mentioned). Similarly, the diagonal entries of $\mathbf{D}_k$ and $\mathbf{L}_k$ are independent and normally distributed variables of unit variance and zero mean except in the context $\mathrm{MoU} > 1 - 10^{-6}$ in which case $D_{k,22} = D_{k,11} + \eta_k$ and $L_{k,22} = L_{k,11} + \eta_k$. $\eta_k$ being a random variable generated to tune the value of MoU. $\mathbf{M}_k$ and $\mathbf{N}_k$ were simulated using (15) and (16).

### A. Exact HJD case

The first experiment is for the exact HJD case. Note that, in the orthogonal case (where we replaced $\mathbf{A}$ by $\mathrm{orth}(\mathbf{A})$), CO-HJD and RO-HJD achieved the same performance with fast convergence rate (typically less than 7 iterations) and hence the corresponding plot is omitted. We present only the results corresponding to the non-orthogonal case. Fig. 1 illustrates the convergence of the considered non-orthogonal JD methods in a case where an ill-conditioned matrix $\mathbf{A}$ is used ($\mathrm{cond}(\mathbf{A}) > 100$). We observe that H-NOODLES did not converge in the large dimensional case and that CJDi and H-CJDi have the best convergence rates. Similar results (omitted here) were observed for a well-conditioned $\mathbf{A}$ ($\mathrm{cond}(\mathbf{A}) < 5$ for $n = 5$ and $\mathrm{cond}(\mathbf{A}) < 50$ for $n = 50$).

---
[7]In that case, performance index (17) does not converge to zero at the algorithm's convergence [12].

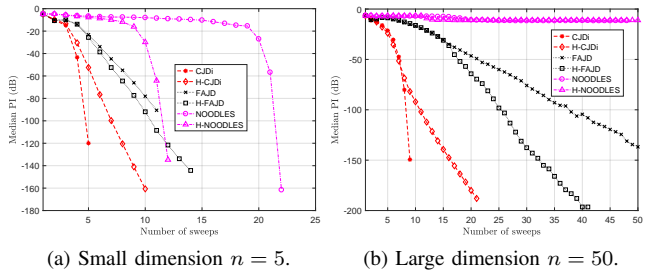

(a) Small dimension $n = 5$.    (b) Large dimension $n = 50$.

Fig. 1. Exact joint diagonalization for small and large dimension. Adverse case with ill-conditioned matrix $\mathbf{A}$ (cond($\mathbf{A}$) $> 100$).

### B. Approximate HJD case

Now, matrices $\mathbf{M}_k$ and $\mathbf{N}_k$ are corrupted by an additive "noise" term with an SNR $= 30$ dB. The results of Fig. 2a (MoU close to 1) show that in the small dimensional case, all non-hybrid algorithms behave similarly and all hybrid algorithms behave similarly too (with a slight advantage for H-CJDi). However, in the large dimensional case, CJDi and H-CJDi are the best. Particularly, we observe a significant gain in favor of the proposed H-CJDi algorithm. Note that in the case where MoU is not close to 1 (not presented here), CJDi and H-CJDi behaved similarly.

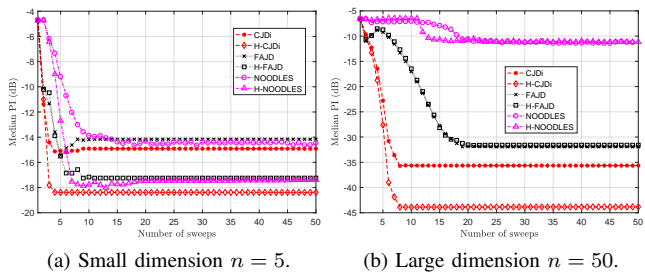

(a) Small dimension $n = 5$.    (b) Large dimension $n = 50$.

Fig. 2. Approximate joint diagonalization for small and large dimension, MoU $\approx 1$ (MoU $> 1 - 10^{-6}$) and SNR $= 30$ dB.

### C. Blind separation of non-circular sources

This last experiment is dedicated to the blind separation of non-circular sources. We consider the model $\mathbf{x}(t) = \mathbf{A}\mathbf{s}(t) + \mathbf{n}(t)$, $t = 0, \ldots, T - 1$ (we chose $T = 1000$ samples) consisting of $m = 5$ instantaneous linear mixtures of $n = 3$ *non-circular*, unit-power, auto-regressive AR(1) sources with AR coefficients $a_1 = 0.95, a_2 = 0.85e^{j\pi/4}$ and $a_3 = 0.7e^{j\pi/6}$, and Gaussian independent innovation processes $\mathbf{o}(t)$ such that $o_i = x_i + jy_i$, $i = 1, \ldots, n$ where vector $[x_i, y_i]^T$ has the following covariance matrix

$$\mathbf{C}_1 = \frac{1}{2}\begin{bmatrix} 1 + \frac{\rho}{\sqrt{2}} & \frac{\rho}{\sqrt{2}} \\ \frac{\rho}{\sqrt{2}} & 1 - \frac{\rho}{\sqrt{2}} \end{bmatrix} \text{ or } \mathbf{C}_2 = \frac{1}{2}\begin{bmatrix} 1 + \frac{\rho}{\sqrt{2}} & 0 \\ 0 & 1 - \frac{\rho}{\sqrt{2}} \end{bmatrix}$$

$0 \leq \rho \leq 1$ being a parameter that controls the non-circularity rate. $\mathbf{C}_1$ is used for all algorithms except ARO-HJD for which $\mathbf{C}_2$ is used since the real and imaginary parts of the source signals need to be independent. The mixtures are corrupted by an additive Gaussian noise. We estimate the mixing matrix $\mathbf{A}$ or equivalently the separation matrix $\mathbf{V}$ through the HJD of $K_1 = 5$ correlation matrices and $K_2 = 5$ pseudo-correlation matrices estimated by appropriate time-averaging over the $T$ observation vectors. The exact noiseless structures of the latter matrices can be shown to correspond to the ones given in (1) and (2) [7]. A pre-whitening is applied first before proceeding to the HJD. Fig. 3 shows the result of comparing orthogonal HJD algorithms for $\rho = 0.1$ and $\rho = 0.9$. We observe that ARO-HJD[8] and CO-HJD perform similarly and that RO-HJD may lead to a performance gain in the case of non-circular signals with high non-circularity rate $\rho$. In Fig. 4a, the noise is spatially white (favorable whitening case) in which case the orthogonal approach CO-HJD gives the best results as compared to SOBI (which uses only the $K_1$ correlation matrices) and to H-CJDi. However, in Fig. 4b we consider a non-favorable case (i.e., a poor whitening condition) where a *decaying* spatial coupling equal to $0.8^\gamma, \gamma = 0, \ldots, m - 1$ exists between the $m$ noise terms. In that case, the orthogonal approach is not the most appropriate anymore and the H-CJDi is the one having the best performance.

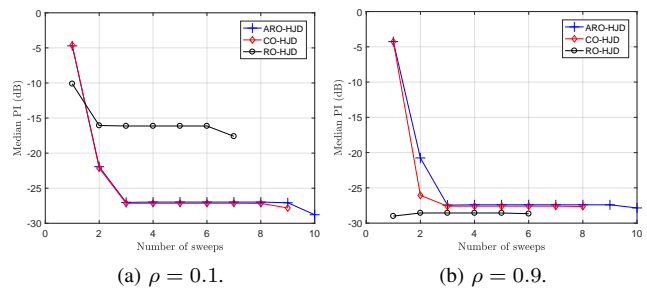

(a) $\rho = 0.1$.    (b) $\rho = 0.9$.

Fig. 3. Comparison of orthogonal JD algorithms for $\rho = 0.1$ and $\rho = 0.9$ with white noise and SNR $= 20$ dB.

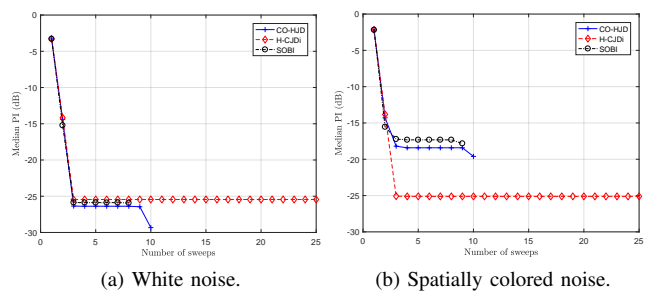

(a) White noise.    (b) Spatially colored noise.

Fig. 4. Blind source separation example for two cases: a (spatially and temporally) white noise and a spatially colored noise both with SNR $= 0$ dB.

## V. CONCLUSION

We introduced new joint diagonalization (JD) algorithms for solving the hybrid case where both Hermitian and transpose congruences are considered. The rationale behind this work is to extend, in this specific context, the most robust and effective orthogonal and non-orthogonal "standard" JD methods, namely SOBI and CJDi. For the former one, several useful and complementary extensions have been proposed in different contexts. As expected, and based on our simulation experiments, the extended algorithms provide improved HJD performance as compared to state-of-the-art methods.

---

[8]Note that for ARO-HJD, the performance index is applied to matrix $\overline{\mathbf{P}} = \overline{\mathbf{V}}^T\overline{\mathbf{A}}$.

# REFERENCES

[1] A. Belouchrani, K. Abed-Meraim, J.-F. Cardoso, and E. Moulines, "A blind source separation technique using second-order statistics," *IEEE Transactions on Signal Processing*, vol. 45, no. 2, pp. 434–444, 1997.

[2] Y. Hua and K. Abed-Meraim, "Techniques of eigenvalues estimation and association," *Digital signal processing*, vol. 7, no. 4, pp. 253–259, 1997.

[3] E. Eidinger and A. Yeredor, "Blind mimo identification using the second characteristic function," *IEEE Transactions on Signal Processing*, vol. 53, no. 11, pp. 4067–4079, 2005.

[4] V. Kuleshov, A. Chaganty, and P. Liang, "Tensor factorization via matrix factorization," in *Artificial Intelligence and Statistics*, 2015, pp. 507–516.

[5] A. Boudjellal, A. Mesloub, K. Abed-Meraim, and A. Belouchrani, "Separation of dependent autoregressive sources using joint matrix diagonalization," *IEEE Signal Processing Letters*, vol. 22, no. 8, pp. 1180–1183, Aug 2015.

[6] T. Adali, P. J. Schreier, and L. L. Scharf, "Complex-valued signal processing: The proper way to deal with impropriety," *IEEE Transactions on Signal Processing*, vol. 59, no. 11, pp. 5101–5125, 2011.

[7] X.-L. Li and T. Adali, "Blind separation of noncircular correlated sources using gaussian entropy rate," *IEEE Transactions on Signal Processing*, vol. 59, no. 6, pp. 2969–2975, 2011.

[8] T. Trainini and E. Moreau, "A coordinate descent algorithm for complex joint diagonalization under Hermitian and transpose congruences," *IEEE Transactions on Signal Processing*, vol. 62, no. 19, pp. 4974–4983, 2014.

[9] A. Yeredor, "Non-orthogonal joint diagonalization in the least-squares sense with application in blind source separation," *IEEE Transactions on Signal Processing*, vol. 50, no. 7, pp. 1545–1553, 2002.

[10] T. Trainini and E. Moreau, "A least squares algorithm for global joint decomposition of complex matrix sets," in *IEEE International Workshop on Computational Advances in Multi-Sensor Adaptive Processing (CAMSAP)*, 2011, pp. 313–316.

[11] L. De Lathauwer and B. De Moor, "On the blind separation of noncircular sources," in *Signal Processing Conference, 2002 11th European*. IEEE, 2002, pp. 1–4.

[12] A. Mesloub, K. Abed-Meraim, and A. Belouchrani, "A new algorithm for complex non-orthogonal joint diagonalization based on shear and Givens rotations," *IEEE Transactions on Signal Processing*, vol. 62, no. 8, pp. 1913–1925, 2014.

[13] J.-F. Cardoso, "On the performance of orthogonal source separation algorithms," in *European Signal Processing Conference (EUSIPCO)*, vol. 94, 1994, pp. 776–779.

[14] J.-F. Cardoso and A. Souloumiac, "Jacobi angles for simultaneous diagonalization," *SIAM journal on matrix analysis and applications*, vol. 17, no. 1, pp. 161–164, 1996.

[15] P. J. Schreier and L. L. Scharf, *Statistical signal processing of complex-valued data: the theory of improper and noncircular signals*. Cambridge University Press, 2010.

[16] A. Belouchrani and W. Ren, "Blind carrier phase tracking with guaranteed global convergence," *IEEE transactions on signal processing*, vol. 45, no. 7, pp. 1889–1894, 1997.

[17] Y. Yao and G. B. Giannakis, "Blind carrier frequency offset estimation in SISO, MIMO, and multiuser OFDM systems," *IEEE Transactions on Communications*, vol. 53, no. 1, pp. 173–183, 2005.

[18] M. Nait-Meziane, K. Abed-Meraim, A.-K. Seghouane, and A. Mesloub, "Supplementary material for the paper: Hybrid Joint Diagonalization Algorithms: Mathematical derivation of CO-HJD, ARO-HJD and H-CJDi," Available at: https://drive.google.com/file/d/1W2crvSx5pg1IFlA6-4onKfelC1kClfeH/view?usp=sharing.

[19] A. Souloumiac, "Nonorthogonal joint diagonalization by combining Givens and hyperbolic rotations," *IEEE Transactions on Signal Processing*, vol. 57, no. 6, pp. 2222–2231, 2009.

[20] X.-L. Li and X.-D. Zhang, "Nonorthogonal joint diagonalization free of degenerate solution," *IEEE Transactions on Signal Processing*, vol. 55, no. 5, pp. 1803–1814, 2007.

[21] B. Afsari, "What can make joint diagonalization difficult?" in *IEEE International Conference on Acoustics, Speech and Signal Processing (ICASSP)*, vol. 3, 2007, pp. 1377–1380.

[22] E. Moreau and O. Macchi, "A one stage self-adaptive algorithm for source separation," in *IEEE International Conference on Acoustics, Speech, and Signal Processing (ICASSP)*, vol. 3, 1994, pp. 49–52.

# Supplementary material for the paper: Hybrid Joint Diagonalization Algorithms
*Mathematical derivation of CO-HJD[*], ARO-HJD[†] and H-CJDi[‡]*

Mohamed Nait-Meziane, Karim Abed-Meraim, Abd-Krim Seghouane, and Ammar Mesloub

We consider two sets of matrices $\mathcal{M} = \{\mathbf{M}_k\}_{1 \leq k \leq K_1}$ and $\mathcal{N} = \{\mathbf{N}_k\}_{1 \leq k \leq K_2}$, diagonal matrices $\mathbf{D}_k$ and $\mathbf{L}_k$, and an unknown full rank matrix $\mathbf{A}$, all in $\mathbb{C}^{n \times n}$ and satisfying $\mathbf{M}_k = \mathbf{A}\mathbf{D}_k\mathbf{A}^H, k = 1, \ldots, K_1$ and $\mathbf{N}_k = \mathbf{A}\mathbf{L}_k\mathbf{A}^T, k = 1, \ldots, K_2$.

## 1 Derivation of CO-HJD

To jointly diagonalize $\mathcal{M}$ and $\mathcal{N}$ we seek a matrix $\mathbf{V}$ that minimizes the following function

$$\mathcal{S}(\mathcal{M}, \mathcal{N}, \mathbf{V}) = \sum_{k=1}^{K_1} \text{off}(\mathbf{V}^H \mathbf{M}_k \mathbf{V}) + \sum_{k=1}^{K_2} \text{off}(\mathbf{V}^H \mathbf{N}_k \mathbf{V}^*) \quad (18)$$

where $\text{off}(\mathbf{X}) = \sum_{1 \leq i \neq j \leq n} |X_{ij}|^2$. In the orthogonal case $\mathbf{V}$ is decomposed as a product of elementary complex Givens rotations $\mathbf{G}_{pq}(\theta, \alpha)$: $\mathbf{V} = \prod_{\#\text{sweeps}} \prod_{1 \leq p < q \leq n} \mathbf{G}_{pq}(\theta, \alpha)$. Hence, $\mathbf{V}$ is updated $n(n-1)/2$ times at each sweep (i.e., going through all index pairs $(p,q)_{1 \leq p < q \leq n}$). $\mathbf{G}_{pq}$[1] is equal to the identity matrix except for its $(p,p)^{\text{th}}$, $(p,q)^{\text{th}}$, $(q,p)^{\text{th}}$, and $(q,q)^{\text{th}}$ entries given by

$$\begin{bmatrix} G_{pp} & G_{pq} \\ G_{qp} & G_{qq} \end{bmatrix} = \begin{bmatrix} \cos(\theta) & -\sin(\theta)e^{-j\alpha} \\ \sin(\theta)e^{j\alpha} & \cos(\theta) \end{bmatrix} = \begin{bmatrix} g_1 & -g_2^* \\ g_2 & g_1 \end{bmatrix}. \quad (19)$$

The parameters of $\mathbf{G}_{pq}$ are computed by minimizing:

$$\mathcal{S}(\mathcal{M}, \mathcal{N}, \mathbf{G}_{pq}) = \mathcal{S}_1(\mathcal{M}, \mathbf{G}_{pq}) + \mathcal{S}_2(\mathcal{N}, \mathbf{G}_{pq}) \quad (20)$$

where $\mathcal{S}_1(\mathcal{M}, \mathbf{G}_{pq}) = \sum_{k=1}^{K_1} \text{off}(\mathbf{G}_{pq}^H \mathbf{M}_k \mathbf{G}_{pq})$ and $\mathcal{S}_2(\mathcal{N}, \mathbf{G}_{pq}) = \sum_{k=1}^{K_2} \text{off}(\mathbf{G}_{pq}^H \mathbf{N}_k \mathbf{G}_{pq}^*)$.

### 1.1 Minimization of $\mathcal{S}_1(\mathcal{M}, \mathbf{G}_{pq})$

Let $\mathbf{M}_k'' = \mathbf{G}_{pq}^H \mathbf{M}_k \mathbf{G}_{pq}$ and $\mathbf{M}_k' = \mathbf{M}_k \mathbf{G}_{pq}$. We have (for $i = 1, \ldots, n$ and $j \neq \{p, q\}$)

$$\begin{cases} M_{k,ip}' = g_1 M_{k,ip} + g_2 M_{k,iq} \\ M_{k,iq}' = -g_2^* M_{k,ip} + g_1 M_{k,iq} \\ M_{k,ij}' = M_{k,ij} \end{cases} \begin{cases} M_{k,pi}'' = g_1 M_{k,pi}' + g_2^* M_{k,qi}' \\ M_{k,qi}'' = -g_2 M_{k,pi}' + g_1 M_{k,qi}' \\ M_{k,ji}'' = M_{k,ji}' \end{cases} \quad (21)$$

---
[*]Complex Orthogonal Hybrid Joint Diagonalization.
[†]Augmented Real Orthogonal Hybrid Joint Diagonalization.
[‡]Hybrid Complex Joint Diagonalization.
[1]We make the dependence of $\mathbf{G}_{pq}$ on $(\theta, \alpha)$ implicit.

Minimizing $\mathcal{S}_1(\mathcal{M}, \mathbf{G}_{pq})$ is equivalent to maximizing[2] $\sum_{k=1}^{K_1} \text{on}(\mathbf{M}_k'')$ where $\text{on}(\mathbf{M}_k'') = \sum_{1 \leq i \leq n} |M_{k,ii}''|^2$, which turns out to be easier to calculate. We have, $\text{on}(\mathbf{M}_k'') = \sum_{\substack{1 \leq i \leq n \\ i \neq \{p,q\}}} |M_{k,ii}|^2 + |M_{k,pp}''|^2 + |M_{k,qq}''|^2$. The first term is independent of $(\theta, \alpha)$. Using the fact that $2(|M_{k,pp}''|^2 + |M_{k,qq}''|^2) = |M_{k,pp}'' - M_{k,qq}''|^2 + |M_{k,pp}'' + M_{k,qq}''|^2$ and the fact that the trace is invariant under a unitary transformation (i.e., $\sum_i M_{k,ii} = \sum_i M_{k,ii}''$) we have $\max_{\theta,\alpha} \sum_{k=1}^{K_1} \text{on}(\mathbf{M}_k'') = \max_{\theta,\alpha} \sum_{k=1}^{K_1} |M_{k,pp}'' - M_{k,qq}''|^2$.

Substituting the relevant indexes in (21) gives $M_{k,pp}'' - M_{k,qq}'' = [|g_1|^2 - |g_2|^2]M_{k,pp} + 2g_1 g_2 M_{k,pq} + 2g_2^* g_1 M_{k,qp} + [|g_2|^2 - |g_1|^2]M_{k,qq}$. Using the trigonometric identities $\sin(2\theta) = 2\sin(\theta)\cos(\theta)$ and $\cos(2\theta) = \cos^2(\theta) - \sin^2(\theta)$ leads to $M_{k,pp}'' - M_{k,qq}'' = \mathbf{e}_{1,k}^T \mathbf{v}$, where $\mathbf{e}_{1,k} = [M_{k,pp} - M_{k,qq}, -(M_{k,pq} + M_{k,qp}), j(M_{k,qp} - M_{k,pq})]^T$ and $\mathbf{v} = [\cos(2\theta), -\sin(2\theta)\cos(\alpha), -\sin(2\theta)\sin(\alpha)]^T$. Hence, $\sum_{k=1}^{K_1} \text{on}(\mathbf{M}_k'') = \sum_{k=1}^{K_1} |\mathbf{e}_{1,k}^T \mathbf{v}|^2 = \mathbf{v}^T \mathbf{E}_1^H \mathbf{E}_1 \mathbf{v}$, where $\mathbf{E}_1^T = [\mathbf{e}_{1,1}, \ldots, \mathbf{e}_{1,K_1}]$ and since $|\mathbf{e}_{1,k}^T \mathbf{v}|^2$ is real-valued, $\mathbf{v}^T \text{Im}(\mathbf{E}_1^H \mathbf{E}_1)\mathbf{v} = 0$ and $\sum_{k=1}^{K_1} \text{on}(\mathbf{M}_k'') = \mathbf{v}^T \text{Re}(\mathbf{E}_1^H \mathbf{E}_1)\mathbf{v}$. Noting that $\mathbf{v}^T \mathbf{v} = 1$ we finally get

$$\min_{\theta,\alpha} \mathcal{S}(\mathcal{M}, \mathbf{G}_{pq}) = \max_{\mathbf{v}} \mathbf{v}^T \text{Re}(\mathbf{E}_1^H \mathbf{E}_1)\mathbf{v} \quad \text{s.t.} \quad \mathbf{v}^T \mathbf{v} = 1. \quad (22)$$

### 1.2 Minimization of $\mathcal{S}_2(\mathcal{N}, \mathbf{G}_{pq})$

Let $\mathbf{N}_k'' = \mathbf{G}_{pq}^H \mathbf{N}_k \mathbf{G}_{pq}^*$ and $\mathbf{N}_k' = \mathbf{N}_k \mathbf{G}_{pq}^*$. We have (for $i = 1, \ldots, n$ and $j \neq \{p, q\}$)

$$\begin{cases} N_{k,ip}' = g_1 N_{k,ip} + g_2^* N_{k,iq} \\ N_{k,iq}' = -g_2 N_{k,ip} + g_1 N_{k,iq} \\ N_{k,ij}' = N_{k,ij} \end{cases} \begin{cases} N_{k,pi}'' = g_1 N_{k,pi}' + g_2^* N_{k,qi}' \\ N_{k,qi}'' = -g_2 N_{k,pi}' + g_1 N_{k,qi}' \\ N_{k,ji}'' = N_{k,ji}' \end{cases} \quad (23)$$

We have $\text{off}(\mathbf{N}_k'') = \sum_{\substack{1 \leq i,j \leq n \\ i,j \neq \{p,q\} \\ i \neq j}} |N_{k,ij}|^2 + |N_{k,pq}''|^2 + |N_{k,qp}''|^2 + \sum_{\substack{1 \leq i \leq n \\ i \neq \{p,q\}}} [|N_{k,ip}''|^2 + |N_{k,iq}''|^2 + |N_{k,pi}''|^2 + |N_{k,qi}''|^2]$. Since the first term is independent of $(\theta, \alpha)$ and $\mathbf{N}_k''$ is complex symmetric ($\mathbf{N}_k = \mathbf{N}_k^T$ and $(\mathbf{N}_k'')^T = (\mathbf{G}_{pq}^H \mathbf{N}_k \mathbf{G}_{pq}^*)^T = \mathbf{G}_{pq}^H \mathbf{N}_k^T \mathbf{G}_{pq}^* = \mathbf{N}_k''$), we have $N_{k,pq}'' = N_{k,qp}''$, $N_{k,ip}'' = N_{k,pi}''$, and $N_{k,iq}'' = N_{k,qi}''$. Hence, $\min_{\theta,\alpha} \sum_{k=1}^{K_2} \text{off}(\mathbf{N}_k'') =$

---
[2]This is because the Frobenius norm is unchanged under unitary transforms.

$\min_{\theta,\alpha} \sum_{k=1}^{K_2} |N''_{k,pq}|^2 + \sum_{\substack{1 \le i \le n \\ i \ne \{p,q\}}} [|N''_{k,ip}|^2 + |N''_{k,iq}|^2]$. Note that $N''_{k,ip}$ and $N''_{k,iq}$ ($i \ne \{p,q\}$) are elements of columns $p$ and $q$ of $\mathbf{N}''_k$, which are only affected by the column transformation of $\mathbf{N}_k$ (i.e., $\mathbf{N}'_k = \mathbf{N}_k \mathbf{G}^*_{pq}$). Hence, $N''_{k,ip} = N'_{k,ip}$ and $N''_{k,iq} = N'_{k,iq}$.

From (23), we have $|N'_{k,ip}|^2 = |g_1|^2|N_{k,ip}|^2 + |g_2|^2|N_{k,iq}|^2 + 2\operatorname{Re}(g_1 N^*_{k,ip} g^*_2 N_{k,iq})$ and $|N'_{k,iq}|^2 = |g_2|^2|N_{k,ip}|^2 + |g_1|^2|N_{k,iq}|^2 - 2\operatorname{Re}(g^*_2 N^*_{k,ip} g_1 N_{k,iq})$. Hence, $|N'_{k,ip}|^2 + |N'_{k,iq}|^2 = (|g_1|^2 + |g_2|^2)|N_{k,ip}|^2 + (|g_1|^2 + |g_2|^2)|N_{k,iq}|^2 = |N_{k,ip}|^2 + |N_{k,iq}|^2$ (since $|g_1|^2 + |g_2|^2 = \cos^2(\theta) + \sin^2(\theta) = 1$). This quantity is independent of $(\theta,\alpha)$ which leads to $\min_{\theta,\alpha} \sum_{k=1}^{K_2} \operatorname{off}(\mathbf{N}''_k) = \min_{\theta,\alpha} \sum_{k=1}^{K_2} |N''_{k,pq}|^2$. Replacing the relevant indexes in (23) we get $N''_{k,pq} = -g_1 g_2 N_{k,pp} + g_1^2 N_{k,pq} - |g_2|^2 N_{k,qp} + g^*_2 g_1 N_{k,qq} = -g_1 g_2 N_{k,pp} + (g_1^2 - |g_2|^2) N_{k,pq} + g^*_2 g_1 N_{k,qq}$. Using the previous trigonometric identities we get $N''_{k,pq} = \mathbf{e}^T_{2,k} \mathbf{v}$, where $\mathbf{e}_{2,k} = \frac{1}{2}[2N_{k,pq}, N_{k,pp} - N_{k,qq}, j(N_{k,pp} + N_{k,qq})]^T$. Hence, $\sum_{k=1}^{K_2} \operatorname{off}(\mathbf{N}''_k) = \sum_{k=1}^{K_2} |\mathbf{e}^T_{2,k} \mathbf{v}|^2 = \mathbf{v}^T \mathbf{E}^H_2 \mathbf{E}_2 \mathbf{v} = \mathbf{v}^T \operatorname{Re}(\mathbf{E}^H_2 \mathbf{E}_2) \mathbf{v}$, where $\mathbf{E}^T_2 = [\mathbf{e}_{2,1}, \ldots, \mathbf{e}_{2,K_2}]$, which leads to

$$\min_{\theta,\alpha} \mathcal{S}_2(\mathcal{N}, \mathbf{G}_{pq}) = \min_{\mathbf{v}} \mathbf{v}^T \operatorname{Re}(\mathbf{E}^H_2 \mathbf{E}_2) \mathbf{v} \quad \text{s.t.} \quad \mathbf{v}^T \mathbf{v} = 1$$
$$= \max_{\mathbf{v}} \left( -\mathbf{v}^T \operatorname{Re}(\mathbf{E}^H_2 \mathbf{E}_2) \mathbf{v} \right) \quad \text{s.t.} \quad \mathbf{v}^T \mathbf{v} = 1. \quad (24)$$

Combining (22) and (24), we finally get

$$\min_{\theta,\alpha} \mathcal{S}(\mathcal{M}, \mathcal{N}, \mathbf{G}_{pq}) = \max_{\mathbf{v}} \mathbf{v}^T \operatorname{Re}(\mathbf{E}^H_1 \mathbf{E}_1 - \mathbf{E}^H_2 \mathbf{E}_2) \mathbf{v}$$
$$\text{s.t.} \quad \mathbf{v}^T \mathbf{v} = 1. \quad (25)$$

Algorithm 1 summarizes the CO-HJD method.

## 2 Derivation of ARO-HJD

Assuming an observation vector $\mathbf{x}(t) = \mathbf{A}\mathbf{s}(t), t = 1, \ldots, T$, $\mathbf{x}(t) \in \mathbb{C}^{m \times 1}$, $\mathbf{A} \in \mathbb{C}^{m \times n}$ and $\mathbf{s}(t) \in \mathbb{C}^{n \times 1}$, we construct the augmented real vector $\overline{\mathbf{x}}(t) = [\operatorname{Re}(\mathbf{x}(t))^T, \operatorname{Im}(\mathbf{x}(t))^T]^T$. We also assume the source signals non-circular with independent real and imaginary parts. The second-order statistics of $\overline{\mathbf{x}}(t)$ have the form $\overline{\mathbf{M}}_k = \overline{\mathbf{A}} \mathbf{D}_k \overline{\mathbf{A}}^T$ where $\mathbf{D}_k$, $k = 1, \cdots, K$ are diagonal matrices and

$$\overline{\mathbf{A}} = \begin{bmatrix} \operatorname{Re}(\mathbf{A}) & -\operatorname{Im}(\mathbf{A}) \\ \operatorname{Im}(\mathbf{A}) & \operatorname{Re}(\mathbf{A}) \end{bmatrix}. \quad (26)$$

To jointly diagonalize the set $\overline{\mathcal{M}} = \{\overline{\mathbf{M}}_k\}_{1 \le k \le K}$, the criterion to minimize is the following

$$\mathcal{L}(\overline{\mathcal{M}}, \overline{\mathbf{V}}) = \sum_{k=1}^{K} \operatorname{off}(\overline{\mathbf{V}}^T \overline{\mathbf{M}}_k \overline{\mathbf{V}}) \quad (27)$$

where $\overline{\mathbf{V}} = \prod_{\#\text{sweeps}} \prod_{p=1}^{n} \mathbf{G}_{p,p+n}(\theta'') \prod_{q=p+1}^{n} \overline{\mathbf{V}}'_{pq}(\theta') \overline{\mathbf{V}}_{pq}(\theta)$ with $\overline{\mathbf{V}}_{pq}(\theta) = \mathbf{G}_{p,q}(\theta) \mathbf{G}_{p+n,q+n}(\theta)$, $\overline{\mathbf{V}}'_{pq}(\theta') =$

---

**Algorithm 1:** Complex Orthogonal Hybrid Diagonalization (CO-HJD) algorithm

**Data:** $\{\mathbf{M}_k \in \mathbb{C}^{n \times n}\}_{1 \le k \le K_1}$, $\{\mathbf{N}_k \in \mathbb{C}^{n \times n}\}_{1 \le k \le K_2}$, $\tau (\ll 1)$.
**Initialization:** $\mathbf{V} \leftarrow \mathbf{I}_{n \times n} + j\mathbf{I}_{n \times n}$.
**while** $\max_{p,q}(|\sin(\theta)|) > \tau$ **do**
  **for** $p = 1$ **to** $n - 1$ **do**
    **for** $q = p + 1$ **to** $n$ **do**
      $\mathbf{v} = [v_1, v_2, v_3]^T \leftarrow$ eigenvector of largest eigenvalue of matrix $\operatorname{Re}(\mathbf{E}^H_1 \mathbf{E}_1 - \mathbf{E}^H_2 \mathbf{E}_2)$;
      Compute elements of $\mathbf{G}_{pq}(\theta, \alpha)$ (see (19)) using:
      $$G_{pp} = G_{qq} \leftarrow \sqrt{(1 + v_1)/2}$$
      $$G_{qp} \leftarrow -(v_2 + jv_3)/(2G_{pp})$$
      $$G_{pq} \leftarrow -G^*_{qp};$$
      $|\sin(\theta)| \leftarrow |G_{qp}|$;
      Update $\{\mathbf{M}_k\}_{1 \le k \le K_1}, \{\mathbf{N}_k\}_{1 \le k \le K_2}$, and $\mathbf{V}$ using:
      $$\mathbf{M}_k \leftarrow \mathbf{G}^H_{pq}(\theta,\alpha) \mathbf{M}_k \mathbf{G}_{pq}(\theta,\alpha)$$
      $$\mathbf{N}_k \leftarrow \mathbf{G}^H_{pq}(\theta,\alpha) \mathbf{N}_k \mathbf{G}^*_{pq}(\theta,\alpha)$$
      $$\mathbf{V} \leftarrow \mathbf{V} \mathbf{G}_{pq}(\theta,\alpha);$$
    **end**
  **end**
**end**

---

$\mathbf{G}_{p,q+n}(\theta') \mathbf{G}_{q,p+n}(\theta')$, and $\mathbf{G}_{p,q}(\theta)$ is a real Givens rotation ($\alpha = 0$).

Minimizing (27) boils down to minimizing at each step (a particular sweep and a particular $p$ and $q$) $\mathcal{L}(\overline{\mathcal{M}}, \overline{\mathbf{V}}_{pq}(\theta))$, $\mathcal{L}(\overline{\mathcal{M}}, \overline{\mathbf{V}}'_{pq}(\theta'))$, and $\mathcal{L}(\overline{\mathcal{M}}, \mathbf{G}_{p,p+n}(\theta''))$.

### 2.1 Minimization of $\mathcal{L}(\overline{\mathcal{M}}, \overline{\mathbf{V}}_{pq}(\theta))$

Let $\overline{\mathbf{M}}''_k = \overline{\mathbf{V}}^T_{pq}(\theta) \overline{\mathbf{M}}_k \overline{\mathbf{V}}_{pq}(\theta) = \overline{\mathbf{V}}^T_{pq}(\theta) \overline{\mathbf{M}}'_k$. We have (for $i = 1, \ldots, 2n$ and $j \ne \{p, q, p+n, q+n\}$)

$$\begin{cases} \overline{M}'_{k,i,p} = \mathsf{c}_\theta \overline{M}_{k,i,p} + \mathsf{s}_\theta \overline{M}_{k,i,q} \\ \overline{M}'_{k,i,q} = -\mathsf{s}_\theta \overline{M}_{k,i,p} + \mathsf{c}_\theta \overline{M}_{k,i,q} \\ \overline{M}'_{k,i,p+n} = \mathsf{c}_\theta \overline{M}_{k,i,p+n} + \mathsf{s}_\theta \overline{M}_{k,i,q+n} \\ \overline{M}'_{k,i,q+n} = -\mathsf{s}_\theta \overline{M}_{k,i,p+n} + \mathsf{c}_\theta \overline{M}_{k,i,q+n} \\ \overline{M}'_{k,i,j} = \overline{M}_{k,i,j} \\ \overline{M}''_{k,p,i} = \mathsf{c}_\theta \overline{M}'_{k,p,i} + \mathsf{s}_\theta \overline{M}'_{k,q,i} \\ \overline{M}''_{k,q,i} = -\mathsf{s}_\theta \overline{M}'_{k,p,i} + \mathsf{c}_\theta \overline{M}'_{k,q,i} \\ \overline{M}''_{k,p+n,i} = \mathsf{c}_\theta \overline{M}'_{k,p+n,i} + \mathsf{s}_\theta \overline{M}'_{k,q+n,i} \\ \overline{M}''_{k,q+n,i} = -\mathsf{s}_\theta \overline{M}'_{k,p+n,i} + \mathsf{c}_\theta \overline{M}'_{k,q+n,i} \\ \overline{M}''_{k,j,i} = \overline{M}'_{k,j,i}. \end{cases} \quad (28)$$

where $\mathsf{c}_\theta = \cos(\theta)$ and $\mathsf{s}_\theta = \sin(\theta)$. Since the Frobenius norm of $\overline{\mathbf{M}}_k$ is unchanged under orthogonal transformation $\overline{\mathbf{V}}_{pq}$, minimizing $\mathcal{L}(\overline{\mathcal{M}}, \overline{\mathbf{V}}_{pq}(\theta))$ is equivalent to maximizing $\sum_{k=1}^{K} \operatorname{on}(\overline{\mathbf{M}}''_k)$ where $\operatorname{on}(\overline{\mathbf{M}}''_k) = \sum_{\substack{1 \le i \le 2n \\ i \ne \{p,q,p+n,q+n\}}} |\overline{M}''_{k,i,i}|^2 + |\overline{M}''_{k,p,p}|^2 + |\overline{M}''_{k,q,q}|^2 + |\overline{M}''_{k,p+n,p+n}|^2 + |\overline{M}''_{k,q+n,q+n}|^2$. The

first term is independent of $\theta$ hence we only need to consider the remaining terms. Using (28) and the trigonometric identities $\sin(2\theta) = 2\sin(\theta)\cos(\theta)$, $\cos^2(\theta) = \frac{1+\cos(2\theta)}{2}$ and $\sin^2(\theta) = \frac{1-\cos(2\theta)}{2}$ we find that $\overline{M}''_{k,p,p} = \sin(2\theta)(\frac{\overline{M}_{k,p,q}+\overline{M}_{k,q,p}}{2}) + \cos(2\theta)(\frac{\overline{M}_{k,p,p}-\overline{M}_{k,q,q}}{2}) + (\frac{\overline{M}_{k,p,p}+\overline{M}_{k,q,q}}{2})$ and $\overline{M}''_{k,q,q} = -\sin(2\theta)(\frac{\overline{M}_{k,p,q}+\overline{M}_{k,q,p}}{2}) + \cos(2\theta)(\frac{\overline{M}_{k,q,q}-\overline{M}_{k,p,p}}{2}) + (\frac{\overline{M}_{k,p,p}+\overline{M}_{k,q,q}}{2})$. Letting $\overline{\mathbf{v}} = [\cos(2\theta), -\sin(2\theta)]^T$, $\mathbf{f}_{1,k} = [\overline{M}_{k,q,q} - \overline{M}_{k,p,p}, \overline{M}_{k,p,q} + \overline{M}_{k,q,p}]^T$, and $c_1 = \overline{M}_{k,p,p} + \overline{M}_{k,q,q}$ we can write $\overline{M}''_{k,p,p} = \frac{1}{2}(-\overline{\mathbf{v}}^T \mathbf{f}_{1,k} + c_1)$ and $\overline{M}''_{k,q,q} = \frac{1}{2}(\overline{\mathbf{v}}^T \mathbf{f}_{1,k} + c_1)$. Hence, $|\overline{M}''_{k,p,p}|^2 + |\overline{M}''_{k,q,q}|^2 = \frac{1}{2}(|\mathbf{f}^T_{1,k}\overline{\mathbf{v}}|^2 + c_1)$. Since $c_1$ is a constant it does not affect the maximization. A similar result is obtained for $|\overline{M}''_{k,p+n,p+n}|^2 + |\overline{M}''_{k,q+n,q+n}|^2$. This gives $\max_\theta \sum_{k=1}^K \text{on}(\overline{\mathbf{M}}''_k) = \max_{\overline{\mathbf{v}}} \sum_{k=1}^K |\mathbf{f}^T_{1,k}\overline{\mathbf{v}}|^2 + |\mathbf{f}^T_{2,k}\overline{\mathbf{v}}|^2$ where $\mathbf{f}_{2,k} = [\overline{M}_{k,q+n,q+n} - \overline{M}_{k,p+n,p+n}, \overline{M}_{k,p+n,q+n} + \overline{M}_{k,q+n,p+n}]^T$. Letting $\mathbf{F}^T_i = [\mathbf{f}_{i,1}, \ldots, \mathbf{f}_{i,K}]$, we get

$$\min_\theta \mathcal{L}(\overline{\mathbf{M}}, \overline{\mathbf{V}}_{pq}(\theta)) = \max_{\overline{\mathbf{v}}} \overline{\mathbf{v}}^T(\mathbf{F}^T_1 \mathbf{F}_1 + \mathbf{F}^T_2 \mathbf{F}_2)\overline{\mathbf{v}}$$
$$\text{s.t.} \quad \overline{\mathbf{v}}^T \overline{\mathbf{v}} = 1 \ . \quad (29)$$

The solution is the principal eigenvector of $\mathbf{F}^T_1 \mathbf{F}_1 + \mathbf{F}^T_2 \mathbf{F}_2$.

## 2.2 Minimization of $\mathcal{L}(\overline{\mathcal{M}}, \overline{\mathbf{V}}'_{pq}(\theta'))$ and $\mathcal{L}(\overline{\mathcal{M}}, \mathbf{G}_{p,p+n}(\theta''))$

A similar derivation to the one presented in section 2.1 leads to the following results:

$$\min_{\theta'} \mathcal{L}(\overline{\mathbf{M}}, \overline{\mathbf{V}}'_{pq}(\theta')) = \max_{\overline{\mathbf{v}}'} \overline{\mathbf{v}}'^T (\mathbf{F}^T_3 \mathbf{F}_3 + \mathbf{F}^T_4 \mathbf{F}_4)\overline{\mathbf{v}}'$$
$$\text{s.t.} \quad \overline{\mathbf{v}}'^T \overline{\mathbf{v}}' = 1 \ , \quad (30)$$

$$\min_{\theta''} \mathcal{L}(\overline{\mathbf{M}}, \mathbf{G}_{p,p+n}(\theta'')) = \max_{\overline{\mathbf{v}}''} \overline{\mathbf{v}}''^T \mathbf{F}^T_5 \mathbf{F}_5 \overline{\mathbf{v}}''$$
$$\text{s.t.} \quad \overline{\mathbf{v}}''^T \overline{\mathbf{v}}'' = 1 \quad (31)$$

where

$\mathbf{f}_{3,k} = [\overline{M}_{k,q+n,q+n} - \overline{M}_{k,p,p}, \overline{M}_{k,p,q+n} + \overline{M}_{k,q+n,p}]^T$,
$\mathbf{f}_{4,k} = [\overline{M}_{k,p+n,p+n} - \overline{M}_{k,q,q}, \overline{M}_{k,q,p+n} + \overline{M}_{k,p+n,q}]^T$,
$\mathbf{f}_{5,k} = [\overline{M}_{k,p+n,p+n} - \overline{M}_{k,p,p}, \overline{M}_{k,p,p+n} + \overline{M}_{k,p+n,p}]^T$.

The summary of the ARO-HJD is presented in Algorithm 2.

## 3 Derivation of H-CJDi

Following the derivation of CJDi [1], and in order to be able to replace the generalized rotation by two successive simpler ones (see details below) allowing for closed-form solutions, we need for matrices $\mathbf{M}_k, k = 1, \ldots, K_1$ to be Hermitian and for matrices $\mathbf{N}_k, k = 1, \ldots, K_2$ to be symmetric. Matrices $\mathbf{N}_k$ are symmetric by construction (or

**Algorithm 2:** Augmented Real Orthogonal Hybrid Joint Diagonalization (ARO-HJD) algorithm

**Data:** $\{\overline{\mathbf{M}}_k \in \mathbb{R}^{2n \times 2n}\}_{1 \le k \le K}$, $\tau (\ll 1)$.
**Initialization:** $\overline{\mathbf{V}} \leftarrow \mathbf{I}_{2n \times 2n}$.
**while** $\max_{p,q}(|\sin(\theta)|) > \tau$ **do**
  **for** $p = 1$ to $n$ **do**
    **for** $q = p+1$ to $n$ **do**
      $\overline{\mathbf{v}} = [\overline{v}_1, \overline{v}_2]^T \leftarrow$ principal eigenvector of matrix $\mathbf{Q} = \mathbf{F}^T_1 \mathbf{F}_1 + \mathbf{F}^T_2 \mathbf{F}_2$;
      Compute elements of $\overline{\mathbf{V}}_{p,q}(\theta)$ using:
$$\cos(\theta) \leftarrow \sqrt{(1 + \overline{v}_1)/2}$$
$$\sin(\theta) \leftarrow -\overline{v}_2/(2\cos(\theta)); \quad (32)$$
      Update $\{\overline{\mathbf{M}}_k\}_{1 \le k \le K}$, and $\overline{\mathbf{V}}$ using:
$$\overline{\mathbf{M}}_k \leftarrow \overline{\mathbf{V}}^T_{p,q}(\theta) \overline{\mathbf{M}}_k \overline{\mathbf{V}}_{p,q}(\theta)$$
$$\overline{\mathbf{V}} \leftarrow \overline{\mathbf{V}} \overline{\mathbf{V}}_{p,q}(\theta);$$
      $\overline{\mathbf{v}}' = [\overline{v}'_1, \overline{v}'_2]^T \leftarrow$ principal eigenvector of matrix $\mathbf{Q}' = \mathbf{F}^T_3 \mathbf{F}_3 + \mathbf{F}^T_4 \mathbf{F}_4$;
      Compute elements of $\overline{\mathbf{V}}'_{p,q}(\theta')$ using $\overline{\mathbf{v}}'$ and (32);
      Update $\{\overline{\mathbf{M}}_k\}_{1 \le k \le K}$, and $\overline{\mathbf{V}}$ using:
$$\overline{\mathbf{M}}_k \leftarrow (\overline{\mathbf{V}}'_{p,q}(\theta'))^T \overline{\mathbf{M}}_k \overline{\mathbf{V}}'_{p,q}(\theta')$$
$$\overline{\mathbf{V}} \leftarrow \overline{\mathbf{V}} \overline{\mathbf{V}}'_{p,q}(\theta');$$
    **end**
    $\overline{\mathbf{v}}'' = [\overline{v}''_1, \overline{v}''_2]^T \leftarrow$ principal eigenvector of matrix $\mathbf{Q}'' = \mathbf{F}^T_5 \mathbf{F}_5$;
    Compute elements of $\mathbf{G}_{p,p+n}(\theta'')$ using $\overline{\mathbf{v}}''$ and (32);
    Update $\{\overline{\mathbf{M}}_k\}_{1 \le k \le K}$, and $\overline{\mathbf{V}}$ using:
$$\overline{\mathbf{M}}_k \leftarrow \mathbf{G}^T_{p,p+n}(\theta'') \overline{\mathbf{M}}_k \mathbf{G}_{p,p+n}(\theta'')$$
$$\overline{\mathbf{V}} \leftarrow \overline{\mathbf{V}} \mathbf{G}_{p,p+n}(\theta'');$$
  **end**
**end**

else in the noisy case, one replaces $\mathbf{N}_k$ by $(\mathbf{N}_k + \mathbf{N}^T_k)/2$) and we only need to construct Hermitian matrices from $\mathbf{M}_k$. This is achieved using the following transformation $(k = 1, \ldots, K_1)$

$$\tilde{\mathbf{M}}_{2k-1} = (\mathbf{M}_k + \mathbf{M}^H_k)/2 = \mathbf{A}\operatorname{Re}(\mathbf{D}_k)\mathbf{A}^H$$
$$\tilde{\mathbf{M}}_{2k} = (\mathbf{M}_k - \mathbf{M}^H_k)/(2j) = \mathbf{A}\operatorname{Im}(\mathbf{D}_k)\mathbf{A}^H \ , \quad (33)$$

which gives a set $\tilde{\mathcal{M}} = \{\tilde{\mathbf{M}}_k\}_{1 \le k \le 2K_1}$ of Hermitian matrices embedding all the information contained in matrices $\mathbf{M}_k$.

In the non-orthogonal case $\mathbf{V}$ is decomposed as a product of elementary complex Givens and elementary complex hyperbolic rotations: $\mathbf{V} = \prod_{\#\text{sweeps}} \prod_{1 \le p < q \le n} \mathbf{R}_{pq}(\theta, \alpha, y, \phi)$ where $\mathbf{R}_{pq}(\theta, \alpha, y, \phi) = \mathbf{G}_{pq}(\theta, \alpha)\mathbf{H}_{pq}(y, \phi)$, with $\mathbf{G}_{pq}$ given in (19) and $\mathbf{H}_{pq}(y, \phi)$ equal to the identity matrix except for its $(p,p)^{\text{th}}$, $(p,q)^{\text{th}}$, $(q,p)^{\text{th}}$, and $(q,q)^{\text{th}}$ entries given by:

$$\begin{bmatrix} H_{pp} & H_{pq} \\ H_{qp} & H_{qq} \end{bmatrix} = \begin{bmatrix} \cosh(y) & \sinh(y)e^{-j\phi} \\ \sinh(y)e^{j\phi} & \cosh(y) \end{bmatrix} \ . \quad (34)$$

Motivated by its effectiveness, we follow hereafter the same procedure proposed in [1]. As a matter of fact, the authors showed that applying $\mathbf{R}_{pq}(\theta, \alpha, y, \phi)$ can be replaced by two successive matrices $\mathbf{R}_{pq}^{(0)} = \mathbf{R}_{pq}(\theta, 0, y, 0)$ and $\mathbf{R}_{pq}^{(\frac{\pi}{2})} = \mathbf{R}_{pq}(\theta', \frac{\pi}{2}, y', \frac{\pi}{2})$ which have the advantage of closed-form solution.

As in CO-HJD, at each step, we seek to minimize:

$$\mathcal{C}(\tilde{\mathcal{M}}, \mathcal{N}, \mathbf{R}_{pq}^{(0)}) = \sum_{k=1}^{2K_1} |[(\mathbf{R}_{pq}^{(0)})^H \tilde{\mathbf{M}}_k \mathbf{R}_{pq}^{(0)}]_{pq}|^2$$
$$+ \sum_{k=1}^{K_2} |[(\mathbf{R}_{pq}^{(0)})^H \mathbf{N}_k (\mathbf{R}_{pq}^{(0)})^*]_{pq}|^2 \quad (35)$$

and

$$\mathcal{C}(\tilde{\mathcal{M}}, \mathcal{N}, \mathbf{R}_{pq}^{(\frac{\pi}{2})}) = \sum_{k=1}^{2K_1} |[(\mathbf{R}_{pq}^{(\frac{\pi}{2})})^H \tilde{\mathbf{M}}_k \mathbf{R}_{pq}^{(\frac{\pi}{2})}]_{pq}|^2$$
$$+ \sum_{k=1}^{K_2} |[(\mathbf{R}_{pq}^{(\frac{\pi}{2})})^H \mathbf{N}_k (\mathbf{R}_{pq}^{(\frac{\pi}{2})})^*]_{pq}|^2 . \quad (36)$$

### 3.1 Minimization of $\mathcal{C}(\tilde{\mathcal{M}}, \mathcal{N}, \mathbf{R}_{pq}^{(0)})$

Let $\mathcal{C}_1(\tilde{\mathcal{M}}, \mathbf{R}_{pq}^{(0)}) = \sum_{k=1}^{2K_1} |[\tilde{\mathbf{M}}_k'']_{pq}|^2$ where $\tilde{\mathbf{M}}_k'' = (\mathbf{R}_{pq}^{(0)})^H \tilde{\mathbf{M}}_k \mathbf{R}_{pq}^{(0)}$, $\tilde{\mathbf{M}}_k' = \tilde{\mathbf{M}}_k \mathbf{R}_{pq}^{(0)}$ and

$$\mathbf{R}_{pq}^{(0)} = \begin{bmatrix} c_\theta \, ch_y - s_\theta \, sh_y & c_\theta \, sh_y - s_\theta \, ch_y \\ c_\theta \, sh_y + s_\theta \, ch_y & c_\theta \, ch_y + s_\theta \, sh_y \end{bmatrix} = \begin{bmatrix} r_{11} & r_{12} \\ r_{21} & r_{22} \end{bmatrix} \quad (37)$$

where $c_\theta = \cos(\theta), s_\theta = \sin(\theta), ch_y = \cosh(y)$, and $sh_y = \sinh(y)$. We have (for $i = 1, \ldots, n$ and $j \neq \{p, q\}$)

$$\begin{cases} \tilde{M}_{k,ip}' = r_{11} \tilde{M}_{k,ip} + r_{21} \tilde{M}_{k,iq} \\ \tilde{M}_{k,iq}' = r_{12} \tilde{M}_{k,ip} + r_{22} \tilde{M}_{k,iq} \\ \tilde{M}_{k,ij}' = \tilde{M}_{k,ij} \end{cases} \begin{cases} \tilde{M}_{k,pi}'' = r_{11}^* \tilde{M}_{k,pi}' + r_{21}^* \tilde{M}_{k,qi}' \\ \tilde{M}_{k,qi}'' = r_{12}^* \tilde{M}_{k,pi}' + r_{22}^* \tilde{M}_{k,qi}' \\ \tilde{M}_{k,ji}'' = \tilde{M}_{k,ji}' . \end{cases} \quad (38)$$

Substituting the relevant indexes in (38) we get $\tilde{M}_{k,pq}'' = r_{11}^* r_{12} \tilde{M}_{k,pp} + r_{11}^* r_{22} \tilde{M}_{k,pq} + r_{21}^* r_{12} \tilde{M}_{k,qp} + r_{21}^* r_{22} \tilde{M}_{k,qq}$. Using the previous trigonometric identities and the hyperbolic trigonometric identities $\sinh(2y) = 2 \sinh(y) \cosh(y)$ and $\cosh(2y) = \cosh^2(y) + \sinh^2(y)$, and also using the fact that $\tilde{M}_{k,pq} = \tilde{M}_{k,qp}^*$ we get $\tilde{M}_{k,pq}'' = \mathbf{e}_{3,k}^T \mathbf{w} + j \operatorname{Im}(\tilde{M}_{k,pq})$, where $\mathbf{e}_{3,k} = \frac{1}{2}[\tilde{M}_{k,pp} + \tilde{M}_{k,qq}, \tilde{M}_{k,pp} - \tilde{M}_{k,qq}, 2 \operatorname{Re}(\tilde{M}_{k,pq})]^T$ and $\mathbf{w} = [sh_{2y}, -s_{2\theta} \, ch_{2y}, c_{2\theta} \, ch_{2y}]^T$. This shows that applying $\mathbf{R}_{pq}^{(0)}$ on $\tilde{\mathbf{M}}_k$ modifies only its real part ($\mathbf{e}_{3,k}^T \mathbf{w}$ is real). Hence, $\sum_{k=1}^{2K_1} |[\tilde{\mathbf{M}}_k'']_{pq}|^2 = \sum_{k=1}^{2K_1} [|\mathbf{e}_{3,k}^T \mathbf{w}|^2 + |\operatorname{Im}(\tilde{M}_{k,pq})|^2] = \mathbf{w}^T \operatorname{Re}(\mathbf{E}_3^H \mathbf{E}_3) \mathbf{w} + \sum_{k=1}^{2K_1} |\operatorname{Im}(\tilde{M}_{k,pq})|^2$, where $\mathbf{E}_3^T = [\mathbf{e}_{3,1}, \ldots, \mathbf{e}_{3,2K_1}]$. Note that $\mathbf{w}$ is normalized with respect to the hyperbolic norm (i.e., $-w_1^2 + w_2^2 + w_3^2 = -sh_{2y}^2 + ch_{2y}^2 = 1$). This can be written as $\mathbf{w}^T \mathbf{J} \mathbf{w} = 1$ where $\mathbf{J} = \operatorname{diag}([-1, 1, 1])$. We finally get

$$\min_{\theta, y} \mathcal{C}_1(\tilde{\mathcal{M}}, \mathbf{R}_{pq}^{(0)}) = \min_{\mathbf{w}} \mathbf{w}^T \operatorname{Re}(\mathbf{E}_3^H \mathbf{E}_3) \mathbf{w} \quad \text{s.t.} \quad \mathbf{w}^T \mathbf{J} \mathbf{w} = 1 . \quad (39)$$

Let $\mathcal{C}_2(\mathcal{N}, \mathbf{R}_{pq}^{(0)}) = \sum_{k=1}^{K_2} |[\mathbf{N}_k'']_{pq}|^2$, where $\mathbf{N}_k'' = (\mathbf{R}_{pq}^{(0)})^H \mathbf{N}_k (\mathbf{R}_{pq}^{(0)})^*$, $\mathbf{N}_k' = \mathbf{N}_k (\mathbf{R}_{pq}^{(0)})^*$. Since $\mathbf{R}_{pq}^{(0)}$ is real-valued, we get similar formulas as (38) for elements of $\mathbf{N}_k''$ and $\mathbf{N}_k'$. Likewise, we get $N_{k,pq}'' = r_{11}^* r_{12} N_{k,pp} + r_{11}^* r_{22} N_{k,pq} + r_{21}^* r_{12} N_{k,qp} + r_{21}^* r_{22} N_{k,qq}$. After simplification and using the fact that $N_{k,pq} = N_{k,qp}$ we get $N_{k,pq}'' = \mathbf{e}_{4,k}^T \mathbf{w}$, where $\mathbf{e}_{4,k} = \frac{1}{2}[N_{k,pp} + N_{k,qq}, N_{k,pp} - N_{k,qq}, 2 N_{k,pq}]^T$. Hence, $\sum_{k=1}^{K_2} |[\mathbf{N}_k'']_{pq}|^2 = \sum_{k=1}^{K_2} [|\mathbf{e}_{4,k}^T \mathbf{w}|^2 = \mathbf{w}^T \operatorname{Re}(\mathbf{E}_4^H \mathbf{E}_4) \mathbf{w}$, where $\mathbf{E}_4^T = [\mathbf{e}_{4,1}, \ldots, \mathbf{e}_{4,K_2}]$ and we get as a result

$$\min_{\theta, y} \mathcal{C}_2(\mathcal{N}, \mathbf{R}_{pq}^{(0)}) = \min_{\mathbf{w}} \mathbf{w}^T \operatorname{Re}(\mathbf{E}_4^H \mathbf{E}_4) \mathbf{w} \quad \text{s.t.} \quad \mathbf{w}^T \mathbf{J} \mathbf{w} = 1 . \quad (40)$$

Combining (39) and (40), we finally get

$$\min_{\theta, y} \mathcal{C}(\tilde{\mathcal{M}}, \mathcal{N}, \mathbf{R}_{pq}^{(0)}) = \min_{\mathbf{w}} \mathbf{w}^T \operatorname{Re}(\mathbf{E}_3^H \mathbf{E}_3 + \mathbf{E}_4^H \mathbf{E}_4) \mathbf{w}$$
$$\text{s.t.} \quad \mathbf{w}^T \mathbf{J} \mathbf{w} = 1 . \quad (41)$$

### 3.2 Minimization of $\mathcal{C}(\tilde{\mathcal{M}}, \mathcal{N}, \mathbf{R}_{pq}^{(\frac{\pi}{2})})$

Here, we follow a similar derivation as in section 3.1 using

$$\mathbf{R}_{pq}^{(\frac{\pi}{2})} = \begin{bmatrix} c_{\theta'} \, ch_{y'} - s_{\theta'} \, sh_{y'} & -j(c_{\theta'} \, sh_{y'} - s_{\theta'} \, ch_{y'}) \\ j(c_{\theta'} \, sh_{y'} + s_{\theta'} \, ch_{y'}) & c_{\theta'} \, ch_{y'} + s_{\theta'} \, sh_{y'} \end{bmatrix} \quad (42)$$

where $c_{\theta'} = \cos(\theta'), s_{\theta'} = \sin(\theta'), ch_{y'} = \cosh(y')$, and $sh_{y'} = \sinh(y')$.

After some derivation we get $\tilde{M}_{k,pq}'' = \mathbf{e}_{5,k}^T \mathbf{w}' + \operatorname{Re}(\tilde{M}_{k,pq})$, where $\mathbf{e}_{5,k} = \frac{j}{2}[-(\tilde{M}_{k,pp} + \tilde{M}_{k,qq}), \tilde{M}_{k,qq} - \tilde{M}_{k,pp}, 2 \operatorname{Im}(\tilde{M}_{k,pq})]^T$ and $\mathbf{w}' = [sh_{2y'}, -s_{2\theta'} \, ch_{2y'}, c_{2\theta'} \, ch_{2y'}]^T$. This shows that applying $\mathbf{R}_{pq}^{(\frac{\pi}{2})}$ on $\tilde{\mathbf{M}}_k$ only modifies its imaginary part ($\mathbf{e}_{5,k}^T \mathbf{w}'$ is pure imaginary). Hence, $\sum_{k=1}^{2K_1} |[\tilde{\mathbf{M}}_k'']_{pq}|^2 = \sum_{k=1}^{2K_1} [|\mathbf{e}_{5,k}^T \mathbf{w}'|^2 + |\operatorname{Re}(\tilde{M}_{k,pq})|^2] = \mathbf{w}'^T \operatorname{Re}(\mathbf{E}_5^H \mathbf{E}_5) \mathbf{w}' + \sum_{k=1}^{2K_1} |\operatorname{Re}(\tilde{M}_{k,pq})|^2$, where $\mathbf{E}_5^T = [\mathbf{e}_{5,1}, \ldots, \mathbf{e}_{5,2K_1}]$. This leads to

$$\min_{\theta', y'} \mathcal{C}_1(\tilde{\mathcal{M}}, \mathbf{R}_{pq}^{(\frac{\pi}{2})}) = \min_{\mathbf{w}'} \mathbf{w}'^T \operatorname{Re}(\mathbf{E}_5^H \mathbf{E}_5) \mathbf{w}' \quad \text{s.t.} \quad \mathbf{w}'^T \mathbf{J} \mathbf{w}' = 1 . \quad (43)$$

Likewise, we get $N_{k,pq}'' = \mathbf{e}_{6,k}^T \mathbf{w}'$, where $\mathbf{e}_{6,k} = \frac{j}{2}[N_{k,pp} - N_{k,qq}, N_{k,pp} + N_{k,qq}, -2j N_{k,pq}]^T$. Hence, $\sum_{k=1}^{K_2} |[\mathbf{N}_k'']_{pq}|^2 = \sum_{k=1}^{K_2} [|\mathbf{e}_{6,k}^T \mathbf{w}'|^2 = \mathbf{w}'^T \operatorname{Re}(\mathbf{E}_6^H \mathbf{E}_6) \mathbf{w}'$, where $\mathbf{E}_6^T = [\mathbf{e}_{6,1}, \ldots, \mathbf{e}_{6,K_2}]$. We get as a result

$$\min_{\theta', y'} \mathcal{C}_2(\mathcal{N}, \mathbf{R}_{pq}^{(\frac{\pi}{2})}) = \min_{\mathbf{w}'} \mathbf{w}'^T \operatorname{Re}(\mathbf{E}_6^H \mathbf{E}_6) \mathbf{w}' \quad \text{s.t.} \quad \mathbf{w}'^T \mathbf{J} \mathbf{w}' = 1 . \quad (44)$$

Combining (43) and (44), we finally get

$$\min_{\theta', y'} \mathcal{C}(\tilde{\mathcal{M}}, \mathcal{N}, \mathbf{R}_{pq}^{(\frac{\pi}{2})}) = \min_{\mathbf{w}'} \mathbf{w}'^T \operatorname{Re}(\mathbf{E}_5^H \mathbf{E}_5 + \mathbf{E}_6^H \mathbf{E}_6) \mathbf{w}'$$
$$\text{s.t.} \quad \mathbf{w}'^T \mathbf{J} \mathbf{w}' = 1 . \quad (45)$$

The H-CJDi method is summarized in Algorithm 3.

**Algorithm 3:** Hybrid Complex Joint Diagonalization (H-CJDi) algorithm

**Data:** $\{\mathbf{M}_k \in \mathbb{C}^{n \times n}\}_{1 \leq k \leq K_1}$, $\{\mathbf{N}_k \in \mathbb{C}^{n \times n}\}_{1 \leq k \leq K_2}$, $\tau$ ($\ll 1$).
**Initialization:** $\mathbf{V} \leftarrow \mathbf{I}_{n \times n} + j\mathbf{I}_{n \times n}$, $\{\tilde{\mathbf{M}}_k\}_{1 \leq k \leq 2K_1}$,
$\mathbf{J} = \mathrm{diag}([-1, 1, 1])$.

**while** $\max_{p,q}(|\sin(\theta)|, |\sinh(y)|) > \tau$ **do**
  **for** $p = 1$ **to** $n - 1$ **do**
    **for** $q = p + 1$ **to** $n$ **do**
      $\mathbf{v} = [v_1, v_2, v_3]^T \leftarrow$ generalized eigenvector of median eigenvalue of $(\mathrm{Re}\left(\mathbf{E}_3^H \mathbf{E}_3 + \mathbf{E}_4^H \mathbf{E}_4\right), \mathbf{J})$;
      **if** $v_3 < 0$ **then** $\mathbf{v} \leftarrow -\mathbf{v}$; $\mathbf{v} \leftarrow \mathbf{v}/\sqrt{\mathbf{v}^T \mathbf{J} \mathbf{v}}$;
      Compute elements of $\mathbf{R}_{pq}^{(0)} = \mathbf{R}(\theta, 0, y, 0)_{pq}$ using:

$$\cos(\theta) \leftarrow \frac{1}{\sqrt{2}}\sqrt{1 + \frac{v_3}{\sqrt{1 + v_1^2}}}$$

$$\sin(\theta) \leftarrow \frac{-v_2}{2\cos(\theta)\sqrt{1 + v_1^2}}$$

$$\cosh(y) \leftarrow \frac{1}{\sqrt{2}}\sqrt{1 + \sqrt{1 + v_1^2}}$$

$$\sinh(y) \leftarrow \frac{v_1}{2\cosh(y)}; \qquad (46)$$

      Update $\{\tilde{\mathbf{M}}_k\}_{1 \leq k \leq 2K_1}, \{\mathbf{N}_k\}_{1 \leq k \leq K_2}$, and $\mathbf{V}$ using:

$$\tilde{\mathbf{M}}_k \leftarrow (\mathbf{R}_{pq}^{(0)})^H \tilde{\mathbf{M}}_k \mathbf{R}_{pq}^{(0)}$$
$$\mathbf{N}_k \leftarrow (\mathbf{R}_{pq}^{(0)})^H \mathbf{N}_k (\mathbf{R}_{pq}^{(0)})^* \qquad (47)$$
$$\mathbf{V} \leftarrow \mathbf{V}\mathbf{R}_{pq}^{(0)};$$

      $\mathbf{v}' = [v_1', v_2', v_3']^T \leftarrow$ generalized eigenvector of median eigenvalue of $(\mathrm{Re}\left(\mathbf{E}_5^H \mathbf{E}_5 + \mathbf{E}_6^H \mathbf{E}_6\right), \mathbf{J})$;
      **if** $v_3' < 0$ **then** $\mathbf{v}' \leftarrow -\mathbf{v}'$; $\mathbf{v}' \leftarrow \mathbf{v}'/\sqrt{\mathbf{v}'^T \mathbf{J} \mathbf{v}'}$;
      Compute elements of $\mathbf{R}_{pq}^{(\frac{\pi}{2})} = \mathbf{R}(\theta', \frac{\pi}{2}, y', \frac{\pi}{2})_{pq}$ using Eqs. (46) (replacing $\theta, y, \mathbf{v}$ by $\theta', y', \mathbf{v}'$);
      Update $\{\tilde{\mathbf{M}}_k\}_{1 \leq k \leq 2K_1}, \{\mathbf{N}_k\}_{1 \leq k \leq K_2}$, and $\mathbf{V}$ using Eqs. (47) (replacing $\mathbf{R}_{pq}^{(0)}$ by $\mathbf{R}_{pq}^{(\frac{\pi}{2})}$);
    **end**
  **end**
**end**

# References

[1] A. Mesloub, K. Abed-Meraim, and A. Belouchrani, "A new algorithm for complex non-orthogonal joint diagonalization based on shear and Givens rotations," *IEEE Transactions on Signal Processing*, vol. 62, no. 8, pp. 1913–1925, 2014.